\documentclass{article}
\usepackage{amsmath}                             
\usepackage{amsfonts}
\input psfig.sty
\def\fig{.}

\def\n{\noindent}

\def\T {\frac{T}{2}}
\def\t {\tau}

\def\H {\mathbb{H}}

\begin{document}
 
\baselineskip .7cm

\author{Navin Khaneja \thanks{To whom correspondence may be addressed. Division of Applied Sciences, Harvard University, Cambridge, MA 02138.
Email:navin@hrl.harvard.edu},\ \ Timo Reiss$^{\dagger}$, \ \ Burkhard Luy$^{\dagger}$,\ \  Steffen J. Glaser \thanks{Institute
of Organic Chemistry and Biochemistry II, Technische Universit\"at M\"unchen,
85747 Garching, Germany.  This work was funded by the Fonds der 
Chemischen Industrie and the Deutsche Forschungsgemeinschaft under 
grant Gl 203/4-1.}}

\title{{\bf Optimal Control of Spin Dynamics \\
in the Presence of Relaxation}}

\maketitle
\begin{center}
{\bf Abstract}
\end{center}
\n Experiments in coherent spectroscopy correspond to control of quantum mechanical ensembles
guiding them from initial to final target states. The control
inputs (pulse sequences) that accomplish these transformations 
should be designed to minimize the effects of relaxation and to optimize the
sensitivity of the experiments. For example in nuclear magnetic resonance (NMR) spectroscopy, a
question of fundamental  importance is what is the maximum efficiency of coherence or
polarization transfer between two spins in the presence of relaxation. Furthermore, what is the
optimal pulse sequence which achieves this efficiency? In this letter, we initiate the study of a
class of control systems, which leads to analytical answers to the above questions. Unexpected
gains in sensitivity  are reported for the most commonly used  experiments in 
NMR spectroscopy.
\newpage
\section{Introduction}
The control of quantum ensembles has many applications,
ranging from coherent spectroscopy to quantum information processing. 
In most applications involving control and manipulation of quantum phenomena, the system of
interest is not isolated but interacts with its environment. This leads to the phenomenon of
relaxation, which in practice results
in  signal loss  and ultimately limits the range of applications. Manipulating
quantum systems in a manner that minimizes 
relaxation losses poses  a fundamental challenge of
utmost practical importance. 
A premier example is the control of spin dynamics in nuclear magnetic resonance (NMR)
spectroscopy \cite{Ernst}. In structural biology, NMR spectroscopy plays an important role
because it is the only technique that allows to determine the structure of biological
macro molecules, such as proteins, in aqueous solution. 
In multidimensional NMR experiments,
transfer of coherence between coupled nuclear spins is a  crucial step.
However with increasing size of molecules or molecular complexes, the rotational tumbling of
the molecules becomes slower and leads to increased relaxation losses.
When these relaxation rates become comparable to the spin-spin couplings, the efficiency of coherence transfer is considerably reduced, leading to poor sensitivity and significantly increased measurement times.

With recent theoretical advances, it has become possible 
 to determine upper bounds for the efficiency of arbitrary
coherence transfer steps in the absence of relaxation \cite{Science}. 
However, from a spectroscopist's perspective, some of the most
important practical (and theoretical) problems have so far been unsolved:

(A) What is the
theoretical upper limit for the coherence transfer efficiency in the presence of relaxation?

(B) How can this theoretical limit be reached experimentally?

The above raised questions can be addressed by methods of optimal control theory. The
framework of optimal control theory was developed to solve problems like finding the best way
to steer a rocket such that it reaches the moon e.g. in minimum time or with minimum fuel. Here
we are interested in computing the optimal way to steer a quantum system from some initial
state to a desired final state with minimum relaxation losses.
In this letter we initiate the study of a class of control systems which gives analytical
solutions to the above raised questions.
It is shown that in contrast to
common belief, widely used standard NMR techniques are far from being optimal and surprising
new transfer schemes emerge. 
\section{Optimal control of nuclear spins under relaxation}
The various relaxation mechanisms in NMR spectroscopy have been well studied 
\cite{Ernst, Redfield}. In liquid solutions, the most important relaxation mechanisms are due to
dipole-dipole interaction (DD) and chemical shift anisotropy (CSA), as well as their interference
effects (e.g. DD-CSA cross correlation terms) \cite{Goldman}.
The optimal control methodology presented here is very general and can take into
account arbitrary relaxation
mechanisms. To demonstrate the ideas and basic principles we focus on an isolated pair  of
heteronuclear spins $I$ (e.g.\
$^1$H) and
$S$  (e.g.\
$^{13}$C or $^{15}$N) with a scalar coupling $J$. Both spins are assumed to be on resonance in a
doubly rotating frame and only
dipole-dipole relaxation is considered.  
This case
approximates for example the situation for deuterated and $^{15}$N-labeled proteins in H$_2$O at
moderately high magnetic fields (e.g. 10 Tesla), where $^{1}$H-$^{15}$N spin pairs are isolated and CSA
relaxation is small. In particular, we focus on
slowly tumbling molecules in so called spin diffusion limit \cite{Ernst}. In this case longitudinal relaxation rates  are negligible compared to 
transverse relaxation rates \cite{Ernst}.

For such coupled two-spin systems, the quantum mechanical equation of motion (Liouville-von Neumann equation) for the density operator $\rho$ \cite{Ernst} is given by
\begin{equation}\label{eq:rho_dot} \dot{\rho} = \pi J [-i 2  I_z S_z, \rho] + \pi k [2 I_z S_z, [
2 I_z S_z , \rho]]. \end{equation}
Here $J$ is the scalar coupling constant and $k$ is the transverse relaxation rate. This relaxation
rate
$k$ depends on various physical parameters, such as the gyromagnetic ratios of the spins, the
internuclear distance,  and the correlation time of the molecular tumbling \cite{Ernst}.
In this letter, we address the problem of finding the maximum efficiency
for the transfers 
\begin{equation}\label{eq:antiphase} I_\alpha \rightarrow 2I_\beta S_\gamma\end{equation}
and 
\begin{equation}\label{eq:inphase}  I_\alpha \rightarrow S_\beta, \end{equation}
where $\alpha$, $\beta$, and $\gamma$ can be $x$, $y$ or $z$.
These transfers
are of central importance for two-dimensional NMR spectroscopy and are conventionally
accomplished by the INEPT (Insensitive Nuclei Enhanced by Polarization
Transfer) \cite{INEPT} (see Fig. 1A) and refocused INEPT
\cite{ref_INEPT} pulse sequence elements, respectively.

 \begin{center}
\begin{figure}[h]
\centerline{\psfig{file= \fig/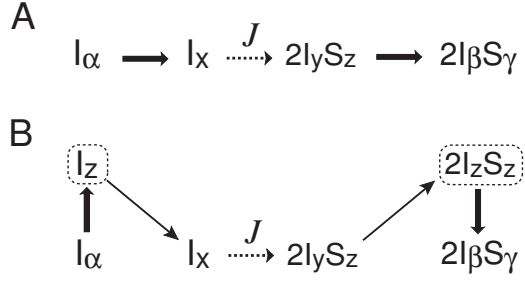,width=2.7 in}}
\caption{\label{fig:scheme} Transfer schemes for (A) INEPT  and (B) ROPE (Relaxation
Optimized Pulse Elements) for the transfer $I_\alpha \rightarrow 2I_\beta S_\gamma$. Thick and
thin arrows represent selective spin rotations by strong and weak rf pulses, respectively.
Dashed arrows represent evolution under $J$ couplings.}
\end{figure}
\end{center} The two heteronuclear spins have  well separated resonance frequencies, allowing for
fast selective manipulation of each  spin on a time-scale determined by the coupling $J$
and the relaxation rate $k$. Hence, in the following it is assumed that any initial Cartesian spin
operator
$I_\alpha$ can be
 transformed
to an operator of the form $ I_x \cos \beta_1 + I_z \sin \beta_1$ by the use of strong,
spin-selective radio frequency (rf) pulses without relaxation losses (see Fig.  2).  
Let $r_1(t)$ represent the magnitude of polarization and in-phase coherence on spin $I$
at any given time $t$, i.e. $r_1^2(t) = \langle I_x\rangle^2 + \langle I_z\rangle^2$, where
$\langle I_{\alpha} \rangle = {\rm trace}\{\rho \ I_{\alpha}  \}$ represents the expectation value of $I_\alpha$.
Using rf fields, we can exactly control  the angle
$\beta_1$ in the term 
$r_1(t) \sin
\beta_1 \ I_z +  r_1(t) \cos \beta_1 \ I_x $.
Hence we can think of $\cos \beta_1$ as a control parameter and denote it by $u_1$ (see Fig. 2).
\begin{center}
\begin{figure}[h]
\centerline{\psfig{file= \fig/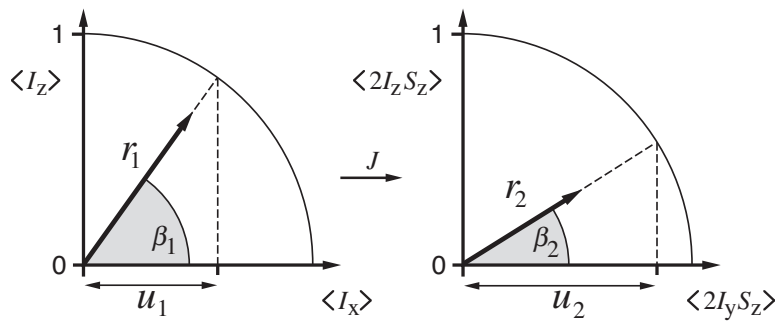 ,width=4in}}
\caption{\label{fig:beta_etc} Representation of the system variables $r_1$, $r_2$, the
angles
$\beta_1$,
$\beta_2$, and of the control parameters
$u_1=\cos
\beta_1$,
$u_2=\cos \beta_2$ in terms of the expectation values $\langle I_x \rangle$, $\langle I_z
\rangle$, $\langle 2 I_y  S_z \rangle$, and $\langle 2 I_z  S_z \rangle$.}
\end{figure}
\end{center} Observe that the operator $I_z$ is invariant under the
evolution equation (1), whereas  $I_x$ evolves under the
$J$ coupling to
$2I_yS_z$ and also relaxes with rate $k$. As the
operator $2I_yS_z$ is produced, it also relaxes with rate $k$. By
use of rf pulses it is possible to rotate the coherence operator
$2I_yS_z$ to $2I_z S_z$, which is protected from relaxation (see Fig. 1 B). Let $r_2$ represent
the total magnitude of the expectation values of these bilinear operators, i.e. $r_2^2(t) = \langle
2I_yS_z\rangle^2 + \langle 2I_zS_z\rangle^2 $. We can control the angle $\beta_2$ in
the term 
$r_2(t)
 \cos \beta_2 \ 2I_yS_z +  r_2 (t) \sin \beta_2 \ 2I_z S_z $
 and we define
 $\cos \beta_2$ as a second control parameter $u_2$ (see Fig. 2). The  evolution of
$r_1(t)$ and $r_2(t)$ under the scalar coupling and relaxation can be expressed as \cite{detail1}
\begin{equation}\label{eq:main.control}\frac{d}{dt} \left[ \begin{array}{c} r_1(t) \\ r_2(t) 
\end{array} \right] = \pi J  \left[ \begin{array}{cc} - \xi u_1^2 & -u_1 u_2 \\ u_1 u_2 & - \xi u_2^2
\end{array}
\right] \left[ \begin{array}{c}
r_1(t) \\ r_2(t) \end{array} \right]. \end{equation} Here 
\begin{equation}\label{eq:XI} \xi = k/J \end{equation} is the relative relaxation rate and measures the relative strength of the relaxation rate $k$ to the spin-spin coupling $J$. 
\begin{center}
\begin{figure}[h]
\centerline{\psfig{file= \fig/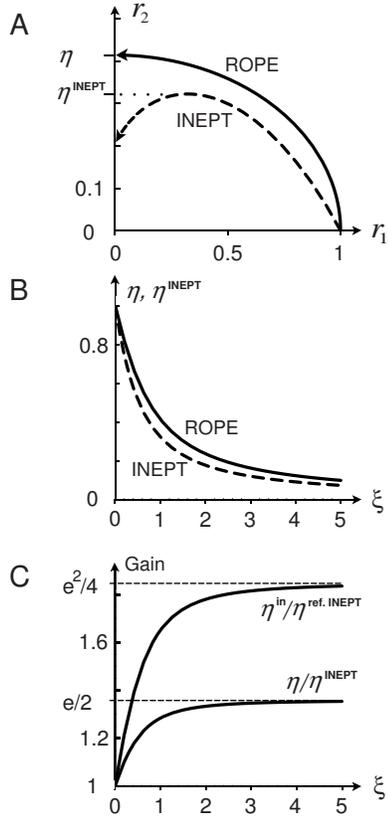 ,width=2
in}}
\caption{\label{fig:dissipation} (A) The dashed curve shows the trajectory 
of the dynamical system (4) when $\xi = 1$
and $u_1(t)$ and $u_2(t)$ are maintained at value 1 (INEPT transfer). The maximum transfer
amplitude
$\eta^{INEPT}$ is reached at $t^{\ast}=(4J)^{-1}$ \cite{detail4}.
The solid curve
represents the trajectory for optimal choice of
$u_1(t)$ and
$u_2(t)$ (ROPE transfer).
(B) Efficiency $\eta^{INEPT}$ of INEPT (dashed curve) and efficiency $\eta$ of ROPE (solid curve) as
a function of the relative relaxation rate $\xi$ for transfer (2). (C) Gain of ROPE transfer efficiency compared to INEPT-type
experiments for transfer (2) ($\eta/\eta^{INEPT}$) and for the in-phase transfer (3)
($\eta^{in}/\eta^{ref INEPT}$).}
\end{figure}
\end{center} The central problem addressed in this paper is the following: Given the dynamical system in equation
(4), how should $u_1(t)$ and $u_2(t)$ be chosen so that starting
from
$r_1(0) = 1$ we achieve the largest value for $r_2$. In spectroscopic applications this would
correspond to the maximum efficiency for the transfer of $I_\alpha$ to $2I_\beta S_\gamma$
(Eq. 2). Observe if $\xi = 0$ (no
relaxation), then by putting $u_1(t) = u_2(t) = 1$, we have $r_2({{1}\over{2J}}) = 1$, i.e. after a
time $t={{1}\over{2J}}$ the operator $I_x$ is completely transferred to
$2I_yS_z$. This is the INEPT transfer element \cite{INEPT}. However if $\xi > 0$, 
it is not the best strategy to keep $u_1(t)$ and $u_2(t)$ both $1$ (as demonstrated
subsequently), see Fig. 3A.
Using principles of optimal control
theory, it is possible to obtain analytical expressions for the largest achievable value of $r_2$
and the optimal values of $u_1(t)$ and $u_2(t)$, see solid curve in Fig. 3A. One of the main
results of the paper is as follows:

\noindent
For the dynamical system in Eq.
(4)
the maximum achievable value of $r_2$ (i.e. the maximum transfer efficiency $\eta$) is given by
\begin{equation}\label{eq:efficiency} 
\eta = \sqrt{1 + \xi^2} - \xi
\end{equation}  and the optimal controls $u_1^\ast(t)$ and
$u_2^\ast(t)$ satisfy the relation
\begin{equation}\label{eq:opt.control} 
{{u_2^{\ast}(t) }\over{u_1^{\ast}(t) }}=\eta {{r_1(t)}\over{r_2(t)}}.
\end{equation} (The optimality of this choice of $u_1$ and $u_2$ can intuitively be seen by the fact that this maximizes the ratio of gain $\delta r_2$ in $r_2$ to loss $\delta r_1$ in $r_1$ for incremental time steps $\delta t$. A more formal proof is given subsequently). The above result implies that throughout the optimal  transfer process, the ratio of 
$\langle 2I_yS_z\rangle(t)$ and $\langle I_x\rangle(t)$ is always maintained at 
\begin{equation}\label{eq:ratio} \frac{\langle 2I_yS_z\rangle(t)}{\langle I_x\rangle(t)}=
\eta, \end{equation} 
(see Fig. 4).

\n Experimentally, the relaxation optimized pulse element (ROPE) which achieves
this optimal efficiency has the following characterization. Starting from the coherence
operator $I_x$ ($r_1=1$, $r_2=0$), this operator is immediately transformed to the polarization
operator $I_z$ (which is protected against relaxation). Then the operator $I_z$ is
gradually rotated towards $I_x$ (which relaxes and also evolves to $2I_yS_z$ under the
coupling term) such that  Eq. (8) is fulfilled for all times. Once $\langle I_z\rangle$
becomes 0,  the operator $2I_yS_z$ is gradually rotated to $2I_zS_z$ (which is also protected
against relaxation), again maintaining the relation of Eq. (8) (see Fig. 4). Finally, $2I_zS_z$ is
rapidly rotated to the target state $2I_\beta
S_\gamma$ (see Fig. 1B). 
 \begin{center}
\begin{figure}[h]
\centerline{\psfig{file= \fig/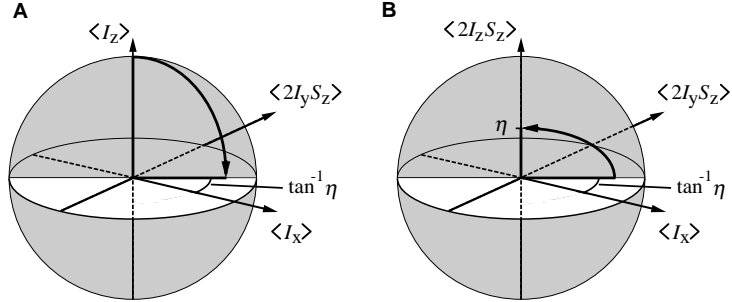,width=3.8in}}
\caption{\label{fig:relation}Schematic representation of the relation (8) to be satisfied by
the optimal trajectory. (A) In the first period during which $\langle I_z\rangle >0$, the
density operator
$\rho$ is restricted to the three-dimensional subspace spanned by the operators $I_x$,  $2I_y
S_z$ and $I_z$.
(B)  In the next period during which $\langle I_z \rangle =0$, the density operator $\rho$
is restricted to the three-dimensional subspace spanned by the operators $I_x$, $2I_y S_z$ and
$2I_z S_z$. The optimal trajectory lies in the plane  which satisfies Eq. (8). }
\end{figure}
\end{center}

It is instructive to compare 
 the optimum coherence transfer efficiency $\eta$ (Eq. 6) for the ROPE transfer (solid curve in
Fig. 3B)  with the maximum transfer efficiency of INEPT which is
 $\eta^{INEPT}=\exp(-\xi \cot^{-1}({\xi})
)\sin(\cot^{-1}({\xi}))$ \cite{detail4} (dashed curve in Fig. 3B).
Figure 3C shows the ratio $\eta/\eta^{INEPT}$ as a function of $\xi$. 
In the limit $\xi \gg 1$ the ratio 
$\eta/\eta^{INEPT}$ approaches e$/2=1.359$. 

\n For the transfer $I_x \rightarrow S_x $ (\ref{eq:inphase}) the operator $I_x$ is first transferred to $2I_zS_y$ which
is then transformed to $S_x$. The optimal  transfer $2 I_zS_y \rightarrow S_x $
is analogous to the optimal transfer $I_x \rightarrow 2 I_z S_y $ and has the same transfer
efficiency. Therefore, the total efficiency for the in-phase to in-phase ROPE transfer is
$\eta^{in}=\eta^2$. Fig. 3C also shows the ratio of this optimal efficiency versus the maximum
efficiency of the refocused INEPT sequence $\eta^{ref. INEPT}=(\eta^{INEPT})^2$. In the limit of
large $\xi$, the ratio approaches e$^2/4=1.847$, i.e. gains of nearly 85 \% are possible using
relaxation optimized pulse elements (ROPE).

\n The proof of the above results is  based on the central tenet of optimal control theory, the
principle of dynamic programming \cite{Bellman}.
In this framework, to find
the optimal way to steer system (4) from the starting point
$(r_1,r_2)=(1,0)$ to the largest possible value
$r_2$, we need to find the best way to steer this system for all
choices of the starting points  $(r_1,r_2)$. Starting from $(r_1,r_2)$, we denote the maximum
achievable value of $r_2$ by $V(r_1, r_2)$, also called the optimal return function for the point
$(r_1, r_2)$. 
The optimal return function for system (4) and optimal control $u_1(r_1, r_2)$ and $u_2(r_1, r_2)$ satisfy the well known Hamilton Jacobi Bellman equation, see \cite{detail15} for details. It can be shown \cite{detail2} that the optimal return function for the control system
(4) is 
\begin{equation}\label{eq:return}
V(r_1, r_2) = \sqrt{\eta^2 r_1^2 + r_2^2}.
\end{equation} and the optimal controls satisfy the equation (\ref{eq:opt.control}).
Evaluating the optimal return function at $(1,0)$, we get 
$V(1,0) = \eta $. Therefore, the maximum transfer efficiency in a spectroscopy experiment
involving transfer of polarization $I_x$ to $2I_zS_x$ is $\eta$
and the optimal controls $u_1$ and $u_2$ satisfy Eq. (7).

\n It is important to note that in the above problem, there is no constraint on the 
the time required to transfer $I_x$ to $2 I_zS_x$. The maximum achievable
efficiency obtained as a solution to the above problem can only be achieved in the limit of very
long transfer times (although most of the efficiency in achieved in finite time).
In practice, it is desirable to reduce the duration of the pulse sequence.
Therefore this raises the question, what is the maximum transfer efficiency  of
$I_\alpha$ to
$2 I_\beta S_\gamma$ in a given finite time $T$.
 This problem can also be explicitly solved (see supplementary material). Here, we describe, the
characteristics of the optimal pulse sequence:
If $T \leq {{\cot^{-1}(2 \xi)}\over{\pi J }}$ then $u_1(t) = u_2(t) = 1$ throughout, i.e. $\beta_1$
and
$\beta_2$ in Fig. 2 are  always kept zero. This solution corresponds to the
INEPT pulse sequence. For  $T > {{\cot^{-1}(2 \xi)}\over{\pi J }}$ the optimal
trajectory has three distinct phases (see Figs. 5 and 6).
\begin{center}
\begin{figure}[h]
\centerline{\psfig{file= \fig/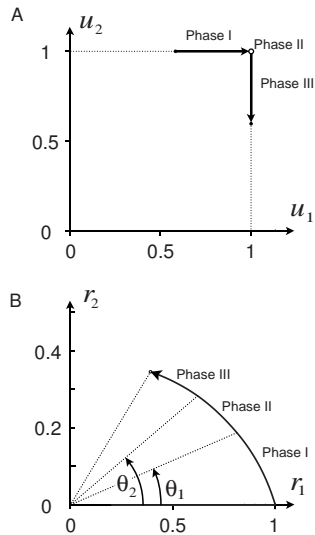,width=1.6in}}
\caption{\label{fig:inept} Phase trajectory of
the controls $u_1$ and $u_2$ (panel A)
 and 
$\vec{r}(t)$ (panel B) for a finite-time ROPE
sequence ($\xi=1$).}
\end{figure}
\end{center}

\begin{center}
\begin{figure}[h]
\centerline{\psfig{file= \fig/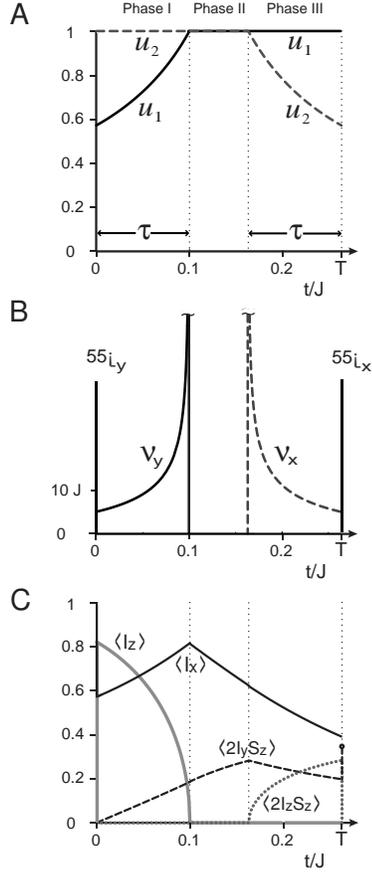,width=1.9in}}
\caption{\label{fig:trajectory} Controls $u_1$ and $u_2$ (panel A), the corresponding rf pulse sequence (panel B) and and the
expectation values  $\langle
I_z\rangle$, $\langle I_x\rangle$,  $\langle
2I_yS_z\rangle$, and  $\langle 2I_zS_z\rangle$ (panel C) are shown
for a finite-time ROPE
sequence ($\xi=1$, $\tau=0.1 J^{-1}$, $T=0.263 J^{-1}$) that optimizes the transfer $I_x
\rightarrow 2 I_y S_z$.
In panel B, the initial hard
$55_{-y}^\circ$ pulse establishes $u_1(0)=0.572$ (see panel A) and the final hard
$55_{-x}^\circ$ pulse completes the transfer. During phase
I and III, the optimal rf amplitudes  $B^{rf}_{x,y}(t)$ are given in frequency units
($\nu_{x,y}(t)=\gamma_I B^{rf}_{x,y}(t)/2 \pi$, where $\gamma_I$ is the gyromagnetic ratio of
spin $I$). During phase II no rf pulses are applied. Approaching phase II (Panel B) the rf amplitude becomes large for a very short time period. This can experimentally be very well approximated by a hard pulse of small flip angle.}
\end{figure}
\end{center} 

There is a
$\tau$ (which is a function of
$T$), such that for 
$0 \leq t \leq \tau$ (phase I), 
$u_2(t) = 1$ and $u_1(t)$ is increased gradually from a value $u_1(0) < 1 \ $
to $u_1(\tau) = 1$ (see supplementary material). Then for time $\tau
\leq t
\leq T-\tau$ (phase II), the optimal control $ u_1(t) = u_2(t) = 1$. Finally for $t \geq T- \tau$
(phase III), we have
$u_1(t) = 1$ and
$u_2(t)$ is decreased from $u_1(T-t) = 1$ to $u_2(T) = u_1(0)$. The optimal control always
satisfies $u_1(t) = u_2(T-t)$, as depicted in Fig. 6A. The
parameter $\tau$, is related to $T$, through the following equation 
\begin{equation}\label{eq:tau}
T = 2 \tau + {{ \theta_2 - \theta_1}\over{\pi J}}\end{equation}
where 
$$\theta_1 = \cot ^{-1}\frac{1 - \kappa(\tau)}{2
\xi \kappa(\tau)}, \ \ 
\theta_2 = \tan ^{-1}\frac{1
- \kappa(\tau)}{2\xi},$$
$$\kappa(\tau) = 1 + 2 \xi^2 - 2 \xi \sqrt{1 + \xi^2}\coth(\pi J \sqrt{1 + \xi^2}\ \tau
+  2 \sinh^{-1}\xi).$$  
At time $\tau$, the optimal trajectory  ($r_1, r_2$) passes from
phase I to II and makes an angle $\theta_1$ with the $r_1$ axis and at time $T-\tau$ the optimal
trajectory passes from phase II to phase III and makes an angle $\theta_2 $ with the $r_1$
axis (see Fig. 5B). The optimal efficiency $\eta_T$ for the finite time $T$ is expressed in terms of these angles as
\begin{equation}\label{eq:opt.u.3} \eta_T = \frac{\exp(\xi(\theta_1 - \theta_2))(1 - \xi \sin 2 \theta_2)}{\sin (\theta_1 + \theta_2)}.\end{equation} In the limit, $T$ goes to infinity $\tau = \frac{T}{2}$ and $\theta_1 = \theta_2 = \tan
^{-1}{\sqrt{1 + \xi^2}  -\xi}$ and $\eta_T$ approaches $\eta$ in (\ref{eq:efficiency}). This corresponds to the unconstrained time case we discussed initially. For the general finite time problem, we can analytically characterize the optimal controls (see Fig. 6A) and the optimal rf pulse elements (see Fig. 6B) \cite{detail3}. Fig. 7 depicts the maximum achievable efficiency as a function of transfer time $T$
for various values of $\xi$.
\begin{center}
\begin{figure}[h]
\centerline{\psfig{file= \fig/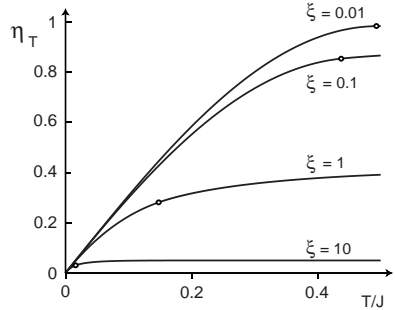,width=2in}}
\caption{\label{fig:eff}
Optimal transfer efficiency $\eta_T$ as a function of the total transfer time T for various values of
$\xi$. The circles indicate the critical time  ${{\cot^{-1}(2 \xi)}\over{\pi J }}$ below which
the rope sequences are identical to the standard INEPT sequence. For times greater than this
critical time, the ROPE sequences are more efficient than INEPT.}
\end{figure}
\end{center}
\section{Conclusions and Outlook} 
In this letter, we initiated the study of  a new class
of control systems 
which arise naturally in  optimal control of quantum mechanical systems in the
presence of relaxation. This made it possible to derive for the first time
upper achievable physical limits on the efficiency of coherence and polarization transfer on two coupled
spins. In this letter, the focus was on the study of 
an  isolated pair  of scalar coupled
heteronuclear spins  under dipole-dipole relaxation in the spin diffusion limit.
For this example a surprising new transfer scheme was found which yields substantial gains (of
up to 85\%) in transfer efficiency.
The results immediately generalize to the case of dipole-dipole and CSA relaxation in the absence
of cross-correlation effects \cite{detail6}. The methods presented here are by no means limited to the case of coupled two spins. These can be generalized for finding relaxation optimized pulse sequences in larger spin systems as commonly encountered in backbone and side chain assignments in protein NMR spectroscopy. Furthermore these methods directly extend to other routinely used experiments like excitation of multiple quantum coherence \cite{Ernst}.    
Some obvious extensions of the methodology presented here are to incorporate  cross-correlation
effects \cite{Goldman} among different relaxation mechanisms and to include  in the design of
pulse sequences additional criteria such as broadbandedness and robustness with respect to
relaxation rates and experimental imperfections. The methods presented here are not restricted to NMR applications
but are broadly applicable to  coherent control of quantum-mechanical phenomena
in the presence of dissipation and decoherence.
The control systems studied in this letter are characterized by the fact that they are linear in the
state of the system and controls can be expressed as polynomial functions of fewer parameters.
Such systems have so far not
received much attention in the optimal control literature due to lack of physical motivation. It is
expected that the study of these systems will foster further developments in the area of system
science and mathematical control theory.

\newpage
\section{Supplementary Material}
We rescale  time to eliminate the factor $\pi J$ in equation (\ref{eq:main.control}). Rewriting (\ref{eq:main.control}) in new time units we get 
\begin{equation}\label{eq:main.control.2}\frac{d}{dt} \left[ \begin{array}{c} r_1(t) \\ r_2(t) 
\end{array} \right] = \left[ \begin{array}{cc} - \xi u_1^2 & -u_1 u_2 \\ u_1 u_2 & - \xi u_2^2
\end{array}
\right] \left[ \begin{array}{c}
r_1(t) \\ r_2(t) \end{array} \right].\end{equation}
In the finite time case, the optimal return function $V(r_1, r_2, t)$ has explicit dependence on time and by definition 
$$ V(r_1, r_2, t) = \max_{u_1, u_2} V(r_1 + \delta t (-\xi u_1^2 r_1 - u_1u_2 r_2), r_2 + \delta t (-\xi u_2^2 r_2 + u_1u_2 r_1), t+ \delta t). $$ Expanding again in powers of $\delta t$ , we obtain the well known Hamilton Jacobi Bellman equation \cite{Bellman}
\begin{equation}\label{eq:jacobi} \frac{\partial V}{\partial t} + \max_{u_1, u_2} \left [ \begin{array}{cc} \frac{\partial V}{\partial r_1} & \frac{\partial V}{\partial r_2} \end{array} \right ] \left [ \begin{array}{cc} -\xi u_1^2 & - u_1u_2 \\  u_1u_2 & - \xi u_2^2 \end{array} \right ]\left [ \begin{array}{c} r_1 \\ r_2 \end{array} \right ] = 0.\end{equation}
As in \cite{detail2}, let $\H = -\lambda_1 r_1  [\xi u_1^2 - (a-b) u_1u_2  + \xi ab u_2^2]$. Then equation (\ref{eq:jacobi}) can be rewritten as 
$$ \frac{\partial V}{\partial t} + \max_{u_1, u_2} \H(u_1, u_2) = 0.$$ For the finite time problem $\max_{u_1, u_2} \H > 0$. This implies $(a-b)^2 > 4 \xi^2 ab$. We consider three separate cases for the problem
\begin{enumerate}
\item {\bf Case I:}  If $(a-b) < 2 \xi$, then the maximum of $\H$ is obtained for $u_2 = 1$ and $u_1 = \frac{a-b}{2 \xi}$.

\item {\bf Case II:} If $(a-b) \geq 2 \xi$ and $\frac{a-b}{ab} \geq 2 \xi$, then the maximum of $\H$ is obtained for $u_1 = 1$ and $u_2 = 1$.

\item {\bf Case III:} If $\frac{a-b}{ab} < 2 \xi$, then the maximum of $\H$ is obtained for $u_1 = 1$ and $u_2 = \frac{a-b}{2 \xi ab}$.
\end{enumerate} 
From equation (\ref{eq:jacobi}), the adjoint variables $(\lambda_1, \lambda_2)= (\frac{\partial V}{\partial r_1}, \frac{\partial
V}{\partial r_2})$ satisfy the equations $\dot{\lambda_1} = - \frac{\partial \H}{\partial r_1}$ and $\dot{\lambda_2} = - \frac{\partial \H}{\partial r_2}$, i.e. 
\begin{equation} \label{eq:adjoint} \frac{d}{dt} \left [ \begin{array}{c} \lambda_1 \\ \lambda_2 \end{array} \right ] = \left [ \begin{array}{cc} \xi u_1^2 & - u_1u_2 \\ u_1u_2 & \xi u_2^2 \end{array} \right ]\left [ \begin{array}{c}\lambda_1 \\ \lambda_2 \end{array} \right ],\end{equation} where $(\lambda_1(T), \lambda_2(T)) = (0,1)$. From equation (\ref{eq:main.control.2}, \ref{eq:adjoint}), we deduce that
$V = \lambda_1 r_1 + \lambda_2 r_2$ is a constant for optimal trajectory and equals the optimal cost $r_2(T) = \lambda_1(0)$. Writing the equation for adjoint variables backward in time,  
let $\sigma = T -t$ then $$ \frac{d}{d\sigma} \left [ \begin{array}{c} \lambda_1 \\ \lambda_2 \end{array} \right ] = \left [ \begin{array}{cc} - \xi u_1^2 & u_1u_2 \\ - u_1u_2 & -\xi u_2^2 \end{array} \right ]\left [ \begin{array}{c}\lambda_1 \\ \lambda_2 \end{array} \right ], $$ where $(\lambda_1(\sigma), \lambda_2(\sigma))_{\sigma = 0} = (0,1)$. Now $u_1(\sigma)$ and $u_2(\sigma)$ should be chosen to maximize $\lambda_1(\sigma)|_{\sigma = T}$. Observe this is exactly the same optimization problem as (\ref{eq:main.control.2}), where the roles of $u_1$ and $u_2$ have been switched. From the symmetry of these two optimization problems, we then have \begin{eqnarray*} 
u_1^{\ast}(t) &=& u_2^{\ast}(T-t) \\
r_1(t) = \lambda_2(T-t) \ &;& \ r_2(t) = \lambda_1(T-t) \\
ab(\T) = 1 &;& V = 2r_1(\T)r_2(\T)
\end{eqnarray*}
Observe from (\ref{eq:main.control.2}, \ref{eq:adjoint}), that $ab(t)$ is monotonically increasing and since $ ab(0) = 0$ and $ ab(\T) = 1$, we have $ab(t) < 1$ for $t < \T$. Therefore $u_2^{\ast}(t) = 1$ for $t < \T$. Since $b(0) = 0$, depending on $a(0)$ we have two cases. 
{\bf Case A} In this case $\frac{a(0)}{2 \xi} \geq 1$. Then we start in the case {II} discussed above and verify that in this case $a-b$ is increasing for $ab <1$. Therefore we stay in this case for all $t \in [0, \T]$ and therefore $u_1^{\ast} = u_2^{\ast}(t) = 1$ for all $t$. Since $b(0)= 0$, we have $b(\T) = \tan T$. Similarly, $$ a(\T) = \frac{a(0) + \tan(\T)}{1 - a(0)\tan(\T)}. $$ If $ab(\T) = 1$ then above equation implies that $\tan(T) \leq \frac{1}{2 \xi}$.

\n {\bf Case B} If $\frac{a(0)}{2 \xi} < 1$, then  $u_1^{\ast}(0) =\frac{a(0)}{2 \xi}$ and the system begins in case {I}. Let $\kappa(t)$ satisfy 
$$ \frac{d \kappa}{dt} = -\frac{\kappa^2 - 2\kappa + 1}{2 \xi} + 2 \xi \kappa, \ \ \kappa(0) = 0 .$$ The solution to this equation is given by $\kappa(t) = 1 + 2 \xi^2 - 2 \xi \sqrt{1 + \xi^2}\coth(\sqrt{1 + \xi^2}t + 2 \beta)$, where $\sinh(\beta) = \xi$. It can be verified that in case I, the optimal trajectory satisfies $\frac{b}{a}(t) = \kappa(t)$. After time $\tau$, $\frac{a-b}{2 \xi}$ becomes equal to $1$ and the system switches to case {II}. Putting $\frac{a-b}{2 \xi} = 1$ and $\frac{b}{a}(t) = \kappa(t)$, we get $ \frac{r_2(\t)}{r_1(\t)} = \frac{2 \xi \kappa(\tau)}{1 - \kappa(\tau)}$ (denote this ratio by $\tan \theta_1$, see Fig 5, Panel B). Then again by symmetry at time $T - \tau$ we have $\frac{1}{2 \xi}(\frac{1}{b} - \frac{1}{a})= 1$ and the system switches from case II to case III. In case III, verify $\frac{b}{a}(t) = \kappa(T-t)$  and the switching to this case occurs at $\tan \theta_2 = \frac{r_2}{r_1} = \frac{1 - \kappa(\tau)}{2 \xi}$. Thus the system spends $T-2 \tau$ in region $II$. Then we have
$$ T - 2 \tau = \tan^{-1} \frac{1- \kappa(\tau)}{2 \xi} -  \tan^{-1} \frac{2 \xi \kappa(\tau)}{1 - \kappa(\tau)}. $$ Thus providing result (\ref{eq:tau}). 

We now derive an explicit expression for $r_2(T)$. For $t \geq T-\tau$, \begin{equation}\label{opt.ret} V(t) = \sqrt{r_2^{2}(t) + \kappa(T-t) r_1^{2}(t)}, \end{equation} is constant along the system trajectories and equals the optimal return function $r_2(T)$. At $t = T - \tau$, we have
$\frac{r_2(T - \tau)}{r_1(T - \tau)} = \tan \theta_2 = \frac{1 - \kappa(\tau)}{2 \xi}$ and therefore from (\ref{opt.ret}), we have \begin{equation}\label{eq:return1}V(T- \tau) = R_1 \sqrt{\sin^2 \theta_2 + \cos^2 \theta_2 - 2 \xi \sin \theta_2 \cos \theta_2}, \end{equation}
where $R_1 = \sqrt{r_1^{2}(t) + r_2^{2}(t)}$ for $t = T - \tau$.  Also note $V(\T) = 2 r_1(\T)r_2(\T)$. At time $t= \T$, we then have $\frac{r_2}{r_1}= \tan(\frac{\theta_1 + \theta_2}{2})$ and therefore 
\begin{equation}\label{eq:return2}
V(\T) = R_2^2 \sin(\theta_1 + \theta_2)
\end{equation}
where $R_2 =\sqrt{r_1^{2}(\T) + r_2^{2}(\T)}$. Note between $\T$ and $T- \tau$, the system evolves under $u_1 = u_2 = 1$. Therefore $R_1 = R_2 \exp(- (\T - \tau))$. Since $V$ is constant, equating (\ref{eq:return1}) and (\ref{eq:return2}), we get equation (\ref{eq:opt.u.3}). 

\begin{thebibliography}{99}

\bibitem{Ernst}
R. R. Ernst, G. Bodenhausen, A. Wokaun, {\it Principles of Nuclear Magnetic Resonance in 
One and Two Dimensions}, (Clarendon Press, Oxford, 1987).

\bibitem{Science}
S. J. Glaser, T. Schulte-Herbr\"uggen, M. Sieveking, O. Schedletzky, N. C. Nielsen, O.
W. S\o rensen, C. Griesinger, {\it Science.} {\bf 208}, 421 (1998).

\bibitem{Redfield}
A. G. Redfield, {\it Adv. Magn. Reson.} {\bf 1}, 1 (1965).

\bibitem{Goldman}
M. Goldman, {\it J. Magn. Reson.} {\bf 60}, 437 (1984).

\bibitem{INEPT}
G. A. Morris, R. Freeman, {\it J. Am. Chem. Soc.} {\bf 101}, 760 (1979).

\bibitem{ref_INEPT}
D. P. Burum, R. R. Ernst, {\it J. Magn. Reson.} {\bf 39}, 163 (1980).

\bibitem{Bellman}
R. Bellman, {\it Dynamic Programming}, (Princeton University Press, Princeton, 1957).

\bibitem{detail1} From equation (\ref{eq:rho_dot}), we have 
$\frac{d\ \langle I_z \rangle (t)}{dt} = 0$, $\frac{d\ \langle I_x \rangle (t)}{dt} = - \pi J\ \langle 2 I_yS_z \rangle(t) - \pi k\ \langle I_x \rangle (t)$, 
 $\frac{d\ \langle 2I_zS_z \rangle(t)}{dt} = 0$ and 
$\frac{d\ \langle  2I_yS_z \rangle(t)}{dt} = \pi J\ \langle I_x \rangle(t) - \pi k\ \langle 2I_ys_z \rangle(t)$. Using the fact that $r_1(t) = \sqrt{{\langle I_z \rangle}^2(t) + {\langle I_x \rangle}^2(t)}$ and  $r_2(t) = \sqrt{{\langle 2I_yS_z \rangle}^2(t) + {\langle 2I_zS_z \rangle}^2(t)}$ and above set of equations, we can write 
\begin{equation} 
\frac{d}{dt} \left[ \begin{array}{c} r_1(t) \\ r_2(t) \end{array} \right] = \pi J \left[ \begin{array}{cc} -
\frac{k}{J} \cos^2 \beta_1(t)  -\cos \beta_1(t) \cos \beta_2(t) \\
\cos \beta_1(t) \cos \beta_2(t)  -\frac{k}{J} \cos^2 \beta_2(t) \end{array} 
\right] \left[ \begin{array}{c} r_1(t) \\ r_2(t) \end{array} \right].
\end{equation}

\bibitem{detail15}If we start at $(r_1, r_2)$, then by making a choice of controls in (\ref{eq:main.control}) and letting the dynamical system evolve, after small time $\delta t$ we can make a transition to all points $(\tilde r_1, \tilde r_2)$, which are related to $(r_1, r_2)$, by 
$$\left [ \begin{array}{c} \tilde r_1 \\ \tilde r_2 \end{array} \right ] = \left[ \begin{array}{c} r_1 \\ r_2 \end{array} \right ] + \delta t \ \pi J \left [ \begin{array}{cc} -\xi u_1^2 & - u_1u_2 \\  u_1u_2 & - \xi u_2^2 \end{array} \right ]\left [ \begin{array}{c} r_1 \\ r_2 \end{array} \right ]; $$ From all points
$(\tilde r_1, \tilde r_2)$ that can be reached by appropriate choice of $(u_1,u_2)$ in small time $\delta t$, we should choose to go to that $(\tilde r_1, \tilde r_2)$ for which $V(\tilde r_1, \tilde r_2)$ is the largest. But now note by definition of $V$ that $V(r_1, r_2) = \max_{\ \tilde r_1, \tilde r_2}V(\tilde r_1, \tilde r_2)$. This can be re-written as 
$$ V(r_1, r_2) = \max_{u_1, u_2} V(r_1 + \delta t (-\xi u_1^2 r_1 - u_1u_2 r_2), r_2 + \delta t (-\xi u_2^2 r_2 + u_1u_2 r_1))$$ for infinitesimal $\delta t$. The right side of the above expression can be expanded (Taylor series expansion) in powers of $\delta t$ and retaining only the terms linear in $\delta t$ (for $\delta t$ approaching zero), we get $$ V(r_1, r_2) = V(r_1, r_2) + \delta t \pi J \max_{u_1,u_2} \left [ \begin{array}{cc} \frac{\partial V}{\partial r_1} & \frac{\partial V}{\partial r_2} \end{array} \right ] \left [ \begin{array}{cc} -\xi u_1^2 & - u_1u_2 \\  u_1u_2 & - \xi u_2^2 \end{array} \right ]\left [ \begin{array}{c} r_1 \\ r_2 \end{array} \right ].$$ Let $\H =  \left [ \begin{array}{cc} \frac{\partial V}{\partial r_1} & \frac{\partial V}{\partial r_2} \end{array} \right ] \left [ \begin{array}{cc} -\xi u^2 & - uv \\  uv & - \xi v^2 \end{array} \right ]\left [ \begin{array}{c} r_1 \\ r_2 \end{array} \right ]$. This equation then reduces to 
\begin{equation}\label{eq: dynamic}
\max_{u_1, u_2} \H(u_1, u_2) = 0.
\end{equation} The optimal control $u_1(r_1, r_2)$ and $u_2(r_1, r_2)$ maximizes the above expression and its maximum value is zero. If we can find a function $V(r_1, r_2)$ , which  satisfies equation (\ref{eq: dynamic}) then finding $(u_1, u_2)$ which satisfy  (\ref{eq: dynamic}) will give us the optimal control to apply in any given state of the dynamical system.

\bibitem{detail2} Let $\H(u_1, u_2)$ be as in [9]. Let $\lambda_1 = \frac{\partial V}{\partial r_1}$, $\lambda_2 = \frac{\partial V}{\partial r_2}$, $a = \frac{\lambda_2}{\lambda_1}$ and $b = \frac{r_2}{r_1}$. Then 
$$ \H = -\lambda_1 r_1  [\xi \ ab\ u_2^2 + (b-a)u_1u_2  + \xi u_1^2]. $$
Observe if $(a-b)\leq 0$, then the only  solution to equations (\ref{eq: dynamic}) is the trivial solution $u_1^{\ast} = u_2^{\ast} = 0$. Therefore $(a-b) > 0$. Also note, when $(a-b)^2 < 4 \xi^2 ab$, the only solution to equation (\ref{eq: dynamic}) is again the trivial solution. Therefore the only case for which (\ref{eq: dynamic}) can be satisfied is if $(b-a)^2 = 4 ab \xi^2$, implying $\sqrt{\frac{b}{a}} = \sqrt{1 + \xi^2} - \xi$. In this regime, maximizing $\H$, we get 
$\frac{u_1^{\ast}}{u_2^{\ast}} = \frac{a-b}{2 \xi}$ implying $\frac{u_1^{\ast}}{u_2^{\ast}} = \frac{b}{\sqrt{1 + \xi^2} - \xi}$.  Integrating equation (\ref{eq:main.control}), for this choice of optimal control, we get that starting from the point $(r_1, r_2)$, the optimal trajectory satisfies that $r_2(t)$ approaches $\sqrt{\eta^2 r_1^2 + r_2^2}$ for large $t$. This is then the desired optimal return function $V(r_1, r_2)$. It can be verified that the optimal return function satisfies equation (\ref{eq: dynamic}). 

\bibitem{detail3}
For $0 \leq t \leq \tau $, the optimal control is given by $$ u_1(t) = \sqrt{\frac{R_1^2
\{ 1 + \cosh(\phi(t)) \}}{(B R_1^2 + 2 A^2 R_2^2) - R_1^2 \cosh(\phi(t))}}, $$
where
 $A = \sinh\phi({{\tau}\over{2}})$,
$B = \cosh \phi(\tau)$ and $\phi(t) = 2 \sinh^{-1}\xi + 2 \pi J t \sqrt{1 + \xi^2}$.
The optimal trajectory crosses from region II to region III at the point
$$ (R_1, R_2) =
(\frac{\eta_T}{\sqrt{\tan^2 \theta_2
+ \kappa(\tau)}} ,  \frac{\eta_T}{\sqrt{1 + \frac{\kappa(\tau)}{\tan^2 \theta_2}}}\ ).$$   (as depicted
in Fig. 5).
For $t > \tau$, we have $u_1(t) = 1$ and $u_2(t) = u_1(T-t)$. The explicit expression for $\nu_{y}$ for phase I in panel B of Fig. 6 in terms of $u_1$ is
$$ \nu_y(t) = \pi J u_1 \{ \xi (1 - \sqrt{1 - u^2}) + \tanh(\frac{\phi}{2})\sqrt{1 + \xi^2}[ 1 + \frac{1 + u_1^2}{\sqrt{1 - u_1^2}}]\ \}, $$ and in phase III,
 $\nu_x(t) = \nu_y(T-t)$. For the transfer $I_z \rightarrow 2I_zS_z$, the flip angle of the initial and final hard pulses (see Fig. 6B) is given by $\sin^{-1}u_1(0)$. For $\xi = 1$ and $T = \frac{0.263}{J}$ we find $u_1(0) = .5716$. The resulting value for initial and final flip angle is $55.138^{\circ}$. 
\bibitem{detail4}
In INEPT,
the efficiency of the transfer $I_x \rightarrow 2I_zS_y $ as a function of transfer time t is
given by
$\eta^{INEPT}(t) = 
\exp(-\pi k t) \sin(\pi J t).$
This efficiency is maximized for a transfer time
$t^{\ast} = \frac{1}{\pi J} \cot^{-1}(\xi)$ and this value is
$\eta^{INEPT}(t^{\ast})=\exp(-\xi \cot^{-1}({\xi})
)\sin(\cot^{-1}({\xi})).$

\bibitem{detail6} In the presence of both CSA and dipole-dipole relaxation (with no cross-correlation effects) the evolution of the density operator takes the form $$\dot{\rho} = \pi J [-i 2  I_z S_z, \rho] + \pi k_1 [2 I_z S_z, [2 I_z S_z , \rho]] + \pi k_2 [I_z, [I_z, \rho]] + \pi k_3 [S_z, [S_z , \rho]] .$$ In this case the operators $I_x$ and $2 I_yS_z$ relax with effective rates $k = k_1 + k_2$ and the operators $2I_zS_y$ and $S_x$ relax with effective rates 
$k' = k_1 + k_3$. Therefore the optimal efficiency of the transfer $I_{\alpha} \rightarrow 2 I_zS_{\beta}$ is $\sqrt{1 + (\frac{k}{J})^2} - \frac{k}{J}$ and the optimal efficiency of the transfer $2 I_zS_{\beta} \rightarrow S_{\alpha}$ is $\sqrt{1 + (\frac{k'}{J})^2} - \frac{k'}{J}$.
\end{thebibliography}
\end{document}